\documentclass{pasj00}
\draft

\begin{document}
\SetRunningHead{Author(s) in page-head}{Running Head}
\Received{2001 June 14}
\Accepted{2002 Jan 1}

\title{Variation of Molecular Cloud Properties across 
the Spiral Arm in M 51}

 


%
 \author{%
   Tomoka \textsc{Tosaki}\altaffilmark{1},
   Takashi \textsc{Hasegawa}\altaffilmark{1},
   Yasuhiro \textsc{Shioya}\altaffilmark{2}, \\
   Nario \textsc{Kuno}\altaffilmark{3}
   and
   Satoki \textsc{Matsushita}\altaffilmark{4}}
 \altaffiltext{1}{Gunma Astronomical Observatory, Nakayama, Takayama, 
                  Agatsuma,Gunma 377-0702}
 \email{tomoka@astron.pref.gunma.jp}
 \altaffiltext{2}{Tohoku University, Aoba, Sendai, Miyagi 980-8578}
 \altaffiltext{3}{Nobeyama Radio Observatory, Minamimaki, Minamisaku, Nagano 
                  384-1805}
 \altaffiltext{4}{Submillimeter Array, Harvard-Smithsonian Center for
                  Astrophysics, P.O.Box 824, Hilo, HI \\ 96721-0824, U.S.A.}

\KeyWords{galaxies: individual(M 51)---  galaxies: ISM --- galaxies: spiral} 

\maketitle

\begin{abstract}

 We present the results of high-resolution $^{13}$CO ($J$~=~1--0) 
mapping observations with the NRO 45m telescope
of the area toward the southern bright arm region of M 51, 
including the galactic center,
in order to study the physical conditions of the 
molecular clouds in the arm and the interarm.
The obtained map shows the central depression of 
the $^{13}$CO ($J$~=~1--0) emission, the circumnuclear ring 
(radius $\sim$ \timeform{10''} -- \timeform{20''}), 
and the spiral arm structure. 
The arm-to-interarm ratio of the $^{13}$CO ($J$~=~1--0) 
integrated intensity is 2 -- 4.
We also have found a feature different from that found in
the $^{12}$CO results. 
For example, the $^{13}$CO distribution shows a depression 
in part of the spiral arm.
The $^{12}$CO/$^{13}$CO ratio spatially varies, 
and shows high values ($\sim$ 20) 
for the interarm and the central region, but low values ($\sim$ 10) for the arm. 
Their values indicate that there is a denser gas in the spiral arm than  
in the interarm.

The distribution of the $^{13}$CO shows a better correspondence with
that of the H$\alpha$ emission than with the $^{12}$CO in the disk region, 
except for the central region.
We found that the $^{13}$CO emission is located on the downstream
side of the $^{12}$CO arm, namely there is an offset between 
the $^{12}$CO and the $^{13}$CO as well as the H$\alpha$ emission.
This suggests that there is a time delay between the accumulation of gas 
caused by the density wave and dense gas formation, accordingly 
star formation.
This time delay is estimated to be $\sim$ 10$^7$ yr based  
on the assumption of galactic rotation derived by 
the rotation curve and the pattern speed of the M 51 spiral pattern. 
It is similar to the growth timescale of a gravitational instability
in the spiral arm of M 51,
suggesting that the gravitational instability plays an important role 
for dense gas formation.

\end{abstract}

\section{Introduction}

One of the most important aims of research is to develop an 
understanding of the formation
mechanism of dense molecular gas in a study of star formation in galaxies,
because there is a clear correlation between the amount of dense gas and
star formation, namely, stars are born in dense gas clouds.
On the other hand, spiral structures, which are a striking feature of 
galaxies, are thought to have a correlation with dense gas formation 
and/or star formation
because the spiral structures are due to density waves that are expected to 
accumulate or compress the molecular gas. 

Recent observational studies at millimeter and submillimeter wavelengths
have revealed that there exists molecular gas with high temperature 
and/or high density in external galaxies, especially at galactic centers.
These molecular gas properties are greatly different from those of our 
Galaxy, such as Giant Molecular Clouds (GMCs) in the Galactic disk.
However, only a few attempts to understand the properties of molecular gas 
have so far been made in a galactic disk that is 
on and between the spiral arms
(\cite{key-GB93a}a; \cite{key-Kuno95}), 
because emission is weaker in the disks than in the galactic centers.

Because $^{13}$CO ($J$~=~1--0) emission is relatively stronger than other 
molecular line emissions, except for $^{12}$CO ($J$~=~1--0),  
many studies have been made of our Galaxy and external galaxies.
$^{13}$CO is optically thin and a tracer of denser gas
($\sim 10^{3-4}$ cm$^{-3}$) than that traced by $^{12}$CO($J$~=~1--0)
($\sim 10^{2-3}$ cm$^{-3}$), which is optically thick 
and surrounds dense molecular gas.
Also, $^{12}$CO is a good tracer of  
molecular gas mass through the conversion factor from
the integrated intensity of $^{12}$CO to $N_{\rm H_2}$
(column density of H$_2$).
Consequently, the ratio of $^{12}$CO($J$~=~1--0)/$^{13}$CO($J$~=~1--0) 
can be used as an indicator of gas density with a 
low or moderate temperature and normal metallicity. 
The $^{12}$CO($J$~=~1--0)/$^{13}$CO($J$~=~1--0) ratios are 
smaller than $\sim$10 for the Galactic plane 
(\cite{key-Sol79}; \cite{key-Pol88}), 
and for the disk of galaxies, e.g. NGC 891 (\cite{key-Saka97}).
On the other hand, the starburst galaxies have higher ratios 
(\cite{key-Aal95}; \cite{key-Kiku98}). 
We must note that the high ratio is seen in the high latitude clouds 
of our Galaxy which are diffuse cold gas (\cite{key-Pol88}).

To study the nature of molecular clouds in the central region, 
the spiral arm and the interarm of galaxies and 
to consider the effect of spiral arms 
on dense gas formation and star formation, we present the results of 
observations using $^{13}$CO($J$~=~1--0) and H$\alpha$ emission 
toward the inner disk of the nearby spiral galaxy M 51.
M 51 is a famous grand-design spiral galaxy and
has been investigated by numerous observations at various wavelengths,
e.g., optical, infrared, radio, and so on.
This galaxy shows two beautiful spiral arms, and
kinematic indications of density waves also have been found within it.
We also note that an oval/bar potential is seen in the central region
of M 51 which has an Active Galactic Nucleus (AGN: (\cite{key-Tera98}).
The parameters of M 51 are summarized in table \ref{tab:1}.

\begin{table}
  \caption{Adopted parameters of M 51.}\label{tab:1}
  \begin{center}
    \begin{tabular}{ll}
 \hline\hline
  Parameter & Value \\
\hline
  Center position (2000.0)$^a$ & 13$^{\rm h}$29$^{\rm m}$52.71$^{\rm s}$ \\
                           & 47$^\circ$11$^\prime$42.6$^{\prime\prime}$\\
  Morphological type$^b$ & Sbc \\
  Systemic Velocity$^c$  & 463 km s$^{-1}$ \\
  Distance$^d$ & 9.6 Mpc \\
  Position angle$^c$ & 170$^\circ$ \\
  Inclination$^c$ & 20$^\circ$  \\
\hline
{\footnotesize a: \citet{key-TH94}.} \\
{\footnotesize b: \citet{key-dV91}.}  \\
{\footnotesize c: \citet{key-Tully74}.} \\
{\footnotesize d: \citet{key-ST74}.}
    \end{tabular}
  \end{center}
\end{table}


\section{Observations}

\subsection{$^{13}$CO observed with the 45m Telescope at Nobeyama Radio 
Observatory (NRO)}

We performed observations of $^{13}$CO($J$~=~1--0) emissions toward the south 
bright arm region of M 51, including the galactic center,
 between 1994 December and 1995 March with the 45-m telescope 
 at Nobeyama Radio Observatory (NRO).
The full width at a half-power beam (FWHP) was 17$^{\prime\prime}$ 
at a rest frequency of $^{13}$CO($J$~=~1--0) (110.2 GHz), which
corresponds to 782 pc at a distance from M 51 of 9.6 Mpc.
The size of the observed region is \timeform{100''} $\times$
\timeform{120''} (4.6 kpc $\times$ 5.5 kpc) and the observing grid 
has an \timeform{11''} spacing (see figure 1).
The absolute pointing accuracy was checked every hour using a SiO maser
source, and was found to be better than 5$^{\prime\prime}$ 
(peak value) throughout the observations.
The main beam efficiency was $\eta_{\rm mb}$ = 0.38 at 110 GHz.
We used the $2 \times 2$ SIS focal-plane array receiver, 
which can simultaneously observe four positions separated on
the sky by 34$^{\prime\prime}$ each (\cite{key-Suna95}).
The typical system noise temperatures, including atmospheric effects
and antenna ohmic losses, were about 400 K.
The 2048-channel wide-band acousto-optical spectrometers (AOS) were used
as receiver backends. 
The frequency resolution and the channel spacing were 
250 kHz and 125 kHz, respectively, and 
the velocity resolution was 0.68 km s$^{-1}$ at 110 GHz.
The total bandwidth was 250 MHz, corresponding to 680  km s$^{-1}$. 
The integration times of each observing point were 80 -- 100 minutes 
in most of the observed region, except for some of the points.
Because those exceptional points were very short ($\sim$ 5 min), 
we did not use them to make the total integrated map (see subsection 3.1).

\subsection{H$\alpha$ observed with 65cm Telescope at Gunma Astronomical
Observatory (GAO)}

\subsubsection{Imaging observation} 

A H$\alpha$ image of M 51 was obtained 
with the 65cm telescope at the Gunma Astronomical observatory (GAO) 
between 2000 December 29 and 2001 January 11. 
Although the best seeing was \timeform{2.5''},
most of the images used for this study had seeing values of  
between \timeform{3''} and \timeform{3.5''}, which were good enough 
for comparison  
with the CO data taken with the NRO-45m telescope. 
A CCD camera assembled by Apogee was attached to the telescope. 
Installed in this camera was a SITe CCD chip of 1024 $times$ 1024 pixels. 
The pixel size of this chip, 24 $\mu$m, corresponds to \timeform{0.6''} 
in the sky and so this camera has a field of view (FOV) of
\timeform{10.6'} square.
This FOV was wide enough not only to cover the entire galaxy pair 
(M 51 and NGC 5195), but to provide an
 area for a sky-level evaluation.
This camera was cooled down to $-40^\circ$C with a water cycle. 
The temperature was stable within 1$^\circ$C each night. 

Two narrow-band filters were used with a nominal bandpass of 20 \AA \  
 centered at 6584 \AA, and 6620 \AA, corresponding to the wavelength 
of H$\alpha$ and continuum light, respectively. 
These filters were designed to have a boxy-shaped transmittance curve, 
so that the nebular emission could be detected evenly over the galaxy
allowing for the rotation of the galaxy.
We must note that the redshift of M 51 puts H$\alpha$ at 6573 \AA. 
This suggests that the edge of these filters have a much higher
sensitivity to the receding side than the approaching side.
Because the observed region of $^{13}$CO corresponds to 
the receding side, we can say that there was little effect on the obtained
H$\alpha$ distribution compared with the $^{13}$CO map.

We recognize that images from this camera were subject to 
the so-called memory effect: an image has traces of stars
that appeared in preceding exposures. 
Mainly to avoid this effect, we drifted the telescope by approximately 
\timeform{30''} for each exposure. 
Because of angular size of the bulge of M 51 is 
much larger than this, we estimated that the memory effect 
due to the bulge was less than the order of 3\% of 
the H$\alpha$ nebular emission,
though the effect would be significantly suppressed by the drift 
of the telescope
and the clipping algorithm in the image reduction.

The exposure time was set at 10 min for all exposures. 
We restricted ourselves to using 6 images for each filter with good 
image quality (less atmospheric extinction, better seeing,  
and, more importantly, less contamination from the memory effect). 
This gives 60 minutes exposure for each filter in total.

\subsubsection{Image reduction}

Reduction was carried out with IRAF in the standard manner.
Dark frames were combined. 
The mean level of the dark current has 
a monotonic decrease due to the memory effect, since it was obtained 
after illuminating the twilight sky to take flat-field images. 
This decrease introduced some ambiguity in the dark-subtraction. 
However, this ambiguity in the dark current was almost flat across the FOV, 
and we assume that local patterns ( i.e. individual H~{\sc ii} regions ) of 
the nebular emission were not badly corrupted.
Flat-fielding was made with combined twilight sky-flat images. 
The flat-field images are not flat by an order of 5\% 
due to inhomogeneous transmittance across the filter. 
It is not easy at all to remove this inhomogeneity, and although the final image
has a residual of a flat-fielding correction, it is of a much lower level 
($\sim 10$\%) than that of the H$\alpha$ emission of typical H~{\sc ii} 
regions. 
Again, this residual is global and would not corrupt the small scale 
patterns of H$\alpha$ emission discussed in this paper.

Hampered by ambiguities in the dark-subtraction and the flat-fielding, 
we made sky-subtraction with a constant value evaluated in the area 
where it is virtually free from light from the galaxies (M 51 and NGC 5195)
 in the broad-band images.
Scaled by the mean flux of 8 stars appearing in images, 
6 images of each filter were combined after positional registration. 
The scaling has typically 5 -- 10\% ambiguity due to the poor photon 
statistics of the stars in the narrow-band images. 
The positional registration is acceptable within \timeform{0.2''}.
Continuum light was subtracted from the H$\alpha$ image, again, 
scaled by the 8 stellar fluxes which appeared in both images.
The stellar fluxes are consistent only within 5\%.

The H$\alpha$ emission in the bright H~{\sc ii} regions has a signal-to-noise ratio 
of better than 10, if the standard deviation in the sky area is adopted 
for the noise source. 
Taking into account the ambiguity in the subtraction and the scaling, 
we estimate that the H$\alpha$ emission may have 20\% ambiguity, 
but the local patterns (i.e., smaller than the flat-field inhomogeneity) 
would have a much better confidence level.

\begin{figure}
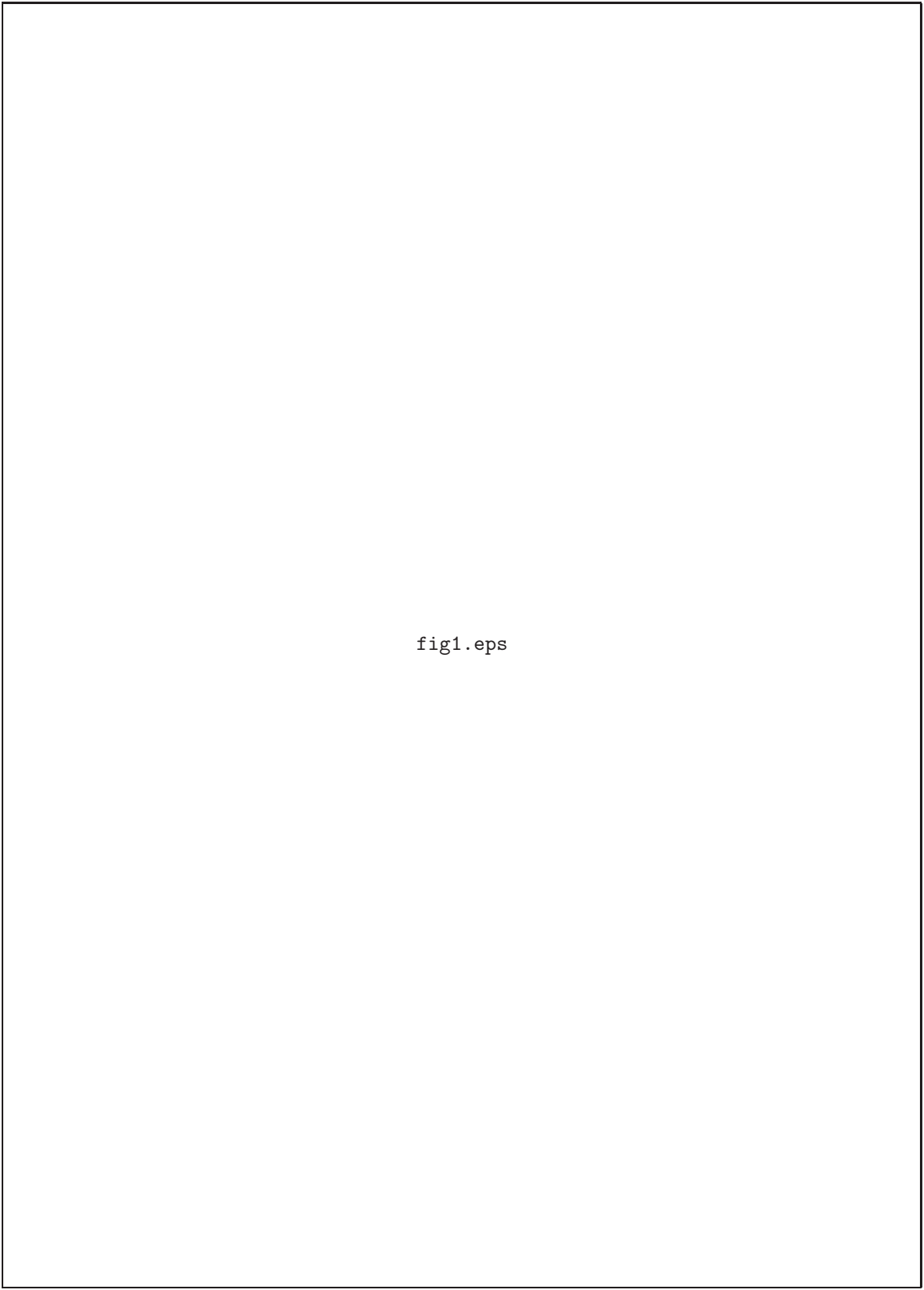

  \begin{center}
    \FigureFile(150mm,210mm){fig1.eps}
  \end{center}
  \caption{H$\alpha$  narrow-band image of M 51(NGC 5194).
  The observed points of $^{13}$CO are indicated by the crosses 
  in the figure. The observing grid uses \timeform{11''} spacing.}
\label{fig:1}
\end{figure}

Figure \ref{fig:1} shows the resultant H$\alpha$ image of M 51.
No efforts were made to correct the extinction 
in the parent galaxy.
This H$\alpha$ emission was largely consistent
with the previous studies. 
A H$\alpha$ image of M 51 was first carefully examined
by \citet{key-Ken89b}, and several studies followed (\cite{key-Rand90}; 
\cite{key-Thilk00}). 
Although quantitative 
a comparison via detailed photometry of individual H~{\sc ii} regions 
is not our aim, 
the remarkable similarity of the emission of bright and faint H~{\sc ii} 
regions 
would support the validity of our observations and data reduction. 
We note that the paucity of H$\alpha$ emission in the center of 
NGC 5195 is consistent with the spectroscopic study (\cite{key-Ho95}), 
giving additional
support to our reduction.

It should also be added that the 3 stars in the observed region
have been used to do {\it absolute} astrometry.
The accuracy has been estimated to a few arcsecs, which is enough to compare
with the $^{12}$CO and $^{13}$CO data.

\section{Results and Discussion}

\subsection{Global Distribution of $^{13}$CO Emission}

\subsubsection{Spectral line profile}

Figure \ref{fig:2} shows a line-profile map of the $^{13}$CO emission 
in the observed region.
As can be seen in this figure, we detected emission in most of the 
observed region.
However, because the emission for a part of the south, west, and north-east 
regions
had poor signal-to-noise, we didn't use these data to make an integrated 
intensity map. 
It should be noted that the emission in the central region of M 51
is very weak, though we found strong emissions surrounding 
the central region, namely the bar-end or the beginning of the spiral arms.
In particular, the west side of the bar-end shows very strong emission.

Figure \ref{fig:2} also shows the detection of emission from the 
interarm as well as strong emission along the spiral arm. 
Figure \ref{fig:2} indicates that a typical peak $T_{\rm mb}$ of the emission 
in the arm is $\sim$ 200 mK and that the west bar-end 
has very strong emission with $T_{\rm mb}$ higher than 300 mK. 
On the other hand, the peak $T_{\rm mb}$ in the interarm is $\sim$
100 mK. 
This means that $T_{\rm mb}$ in the arm is higher than that in the 
interarm by a factor of 2--3. 
We were also able to find a larger velocity dispersion in the arm, 
30 -- 50 km s$^{-1}$, than that in the interarm, $\sim$ 20 km s$^{-1}$. 
The larger velocity width in the arm than in the interarm is 
consistent with previous studies of the $^{12}$CO observations, and
can be explained by the streaming motion at spiral arms caused by  
density wave shock (\cite{key-Kuno97}; \cite{key-GB93a}a).

\begin{figure}
  \begin{center}
    \FigureFile(160mm,150mm){fig2.eps}
  \end{center}
  \caption{Line-profile map of the $^{13}$CO emission spectra 
   with NRO-45m. 
   The velocity resolution of the spectra is 5 km s$^{-1}$.
   The x and y axes of the map are the right ascension and the declination 
   in units of arcsec and (0, 0) in the figure is the galactic center.
   The vertical and horizontal axes of each spectrum are $T_{\rm mb}$ 
   (K) 
   from 0 to 0.4 K and $V_{\rm LSR}$ (km s$^{-1}$) from 300 to 600 
   km s$^{-1}$, 
   respectively.}\label{fig:2}
\end{figure}

\subsubsection{Spiral structure}

\begin{figure}
  \begin{center}
    \FigureFile(150mm,210mm){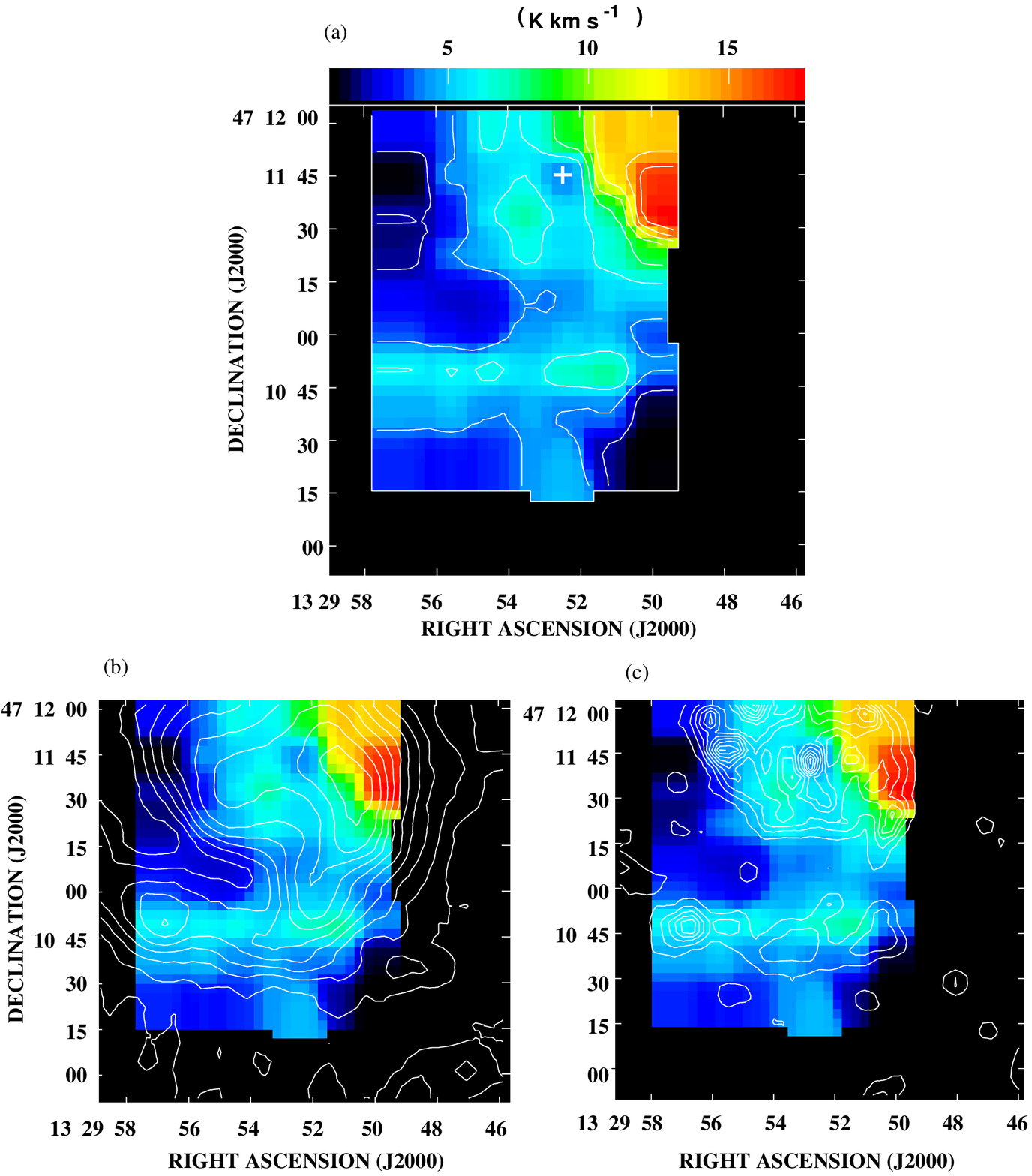}

  \end{center}
  \caption{(a) Total integrated intensity map of $^{13}$CO emission 
   obtained with NRO-45m. The contour interval and the lowest contour is 
   2 K km s$^{-1}$, corresponding to 2$\sigma$. 
   The cross indicates the galactic center of M 51.
   (b) superposition of (a) on the $^{12}$CO emission map 
(contour; \cite{key-Naka94}). 
   The contour interval and the lowest contour is 10 K km s$^{-1}$.
   (c) superposition of (a) on the H$\alpha$ emission map convolved 
to the \timeform{4''} 
   beam
   (contour). The contour interval is 1 in the lowest contour units.}
\label{fig:3}
\end{figure}

The spatial structures of the $^{13}$CO emission are clearly revealed  
in the total integrated intensity map (figure 3). 
Figures \ref{fig:3}a, b, and c show the total integrated map of 
$^{13}$CO emission superposed on a color scale of the $^{13}$CO 
emission, itself,
the $^{12}$CO emission, and the H$\alpha$ emission, respectively.
Figure \ref{fig:3}a shows a depression in the $^{13}$CO($J$~=~1--0) emission 
in the central region of M 51. 
The distribution also has strong concentrations in both bar-ends, 
especially on the west side.
Due to the central depression, we find a ring-like structure surrounding 
the central region.
This depression was pointed out by previous 
interferometric observations of M 51 (\cite{key-Matsu98}).
From figure \ref{fig:2}, 
we have estimated the arm-to-interarm ratio using the 13 and 8 
line profiles of the arm and the interarm, respectively.
We add that the profiles which we cannot distinguish between 
the arm and the interarm due to the location at the edge of the arm
have been excluded.
Their intensities are 4.2 -- 12.6 and 1.3 -- 3.6K km s$^{-1}$ 
for the arm and the interarm, respectively.
The intensities in the arm significantly vary 
while in principle they are increasing with a smaller radius. 
On the other hand, we can not see the large variation in the interarm,
though a slightly larger intensity is seen near 
the beginning of the spiral arm.
Comparing the profiles in the arm with the near ones in the interarm,
we estimate the arm-to-interarm
ratio of the total integrated intensity of the $^{13}$CO ($J$~=~1--0), 
to be 2 -- 4. 
The high ratio can be seen at the beginning of the arm 
or the smaller radius due to the high intensity seen there.
We can also see the arm-to-interarm ratio in the azimuthal distribution 
of the $^{13}$CO in the following figure (figure \ref{fig:8}),
showing the same ratio as that estimated here. 
This indicates that the ratio virtually depends on the intensities in the arm.
Although the ratio seems to be similar to the global ratio of 
the $^{12}$CO ($J$~=~1--0) (\cite{key-Kuno95}),
the arm-to-interarm ratio of the $^{12}$CO decreases 
with a smaller radius toward the galactic center, and consequently 
the arm-to-interarm ratio of the $^{13}$CO 
in the observed region is higher than that of the $^{12}$CO.
We note that the ratio varies spatially in the spiral arm,
showing a high ratio at a smaller radius,
or rather near the beginning of the spiral arm,
as mentioned above.

\begin{figure}
\begin{center}
 \FigureFile(150mm,210mm){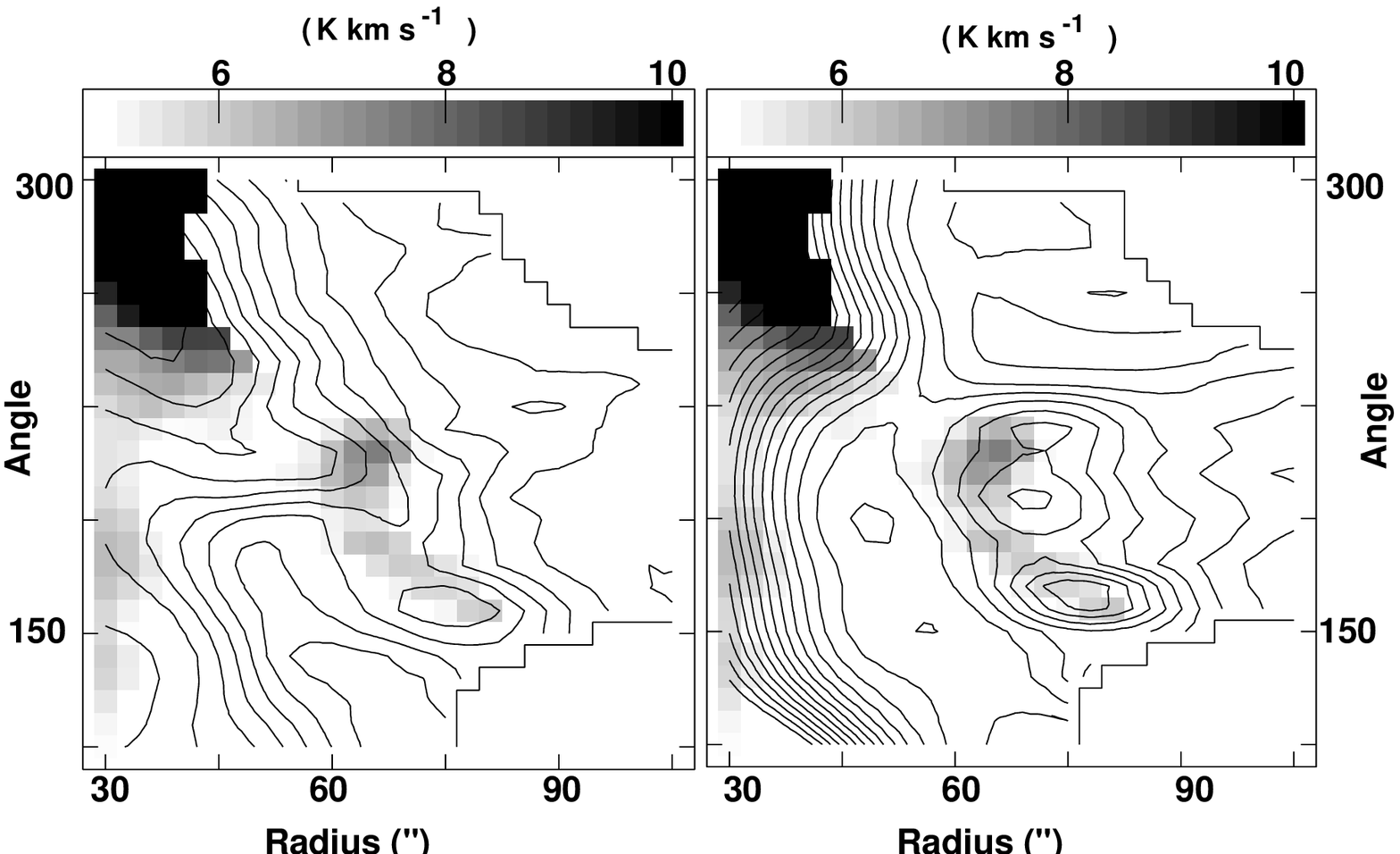}
\end{center}
\caption{$R$ -- $\theta$ plot of the $^{13}$CO intensity (grey scale) 
from $R$ = \timeform{30''} to \timeform{100''}, 
superposed on those of (a) $^{12}$CO emission and 
(b) H$\alpha$ emission (contours), respectively.
The inclination angle has been corrected.
The contour intervals are the same as figure \ref{fig:3}.
The units of the horizontal and the vertical axes are arcsec and degree, 
respectively, and $\theta$ is measured counterclockwise from the minor axis. 
The H$\alpha$ emission is convolved with the \timeform{17''} 
 beam. }
\label{fig:3_2}
\end{figure}

These morphological structures are globally in agreement with
those of the $^{12}$CO emission (figure \ref{fig:3}b).
However, a close look at the distributions of the $^{12}$CO and the $^{13}$CO
will reveal the difference between them.  
First, we can see that there is a depression on the $^{13}$CO spiral arm 
located at (R.A., Decl.) = (\timeform{13h29m51s}, \timeform{47D11'00''}), 
unlike in the $^{12}$CO arm.
In other words, the extension of the spiral arm of the $^{13}$CO from 
the bar-end is shorter than that of the $^{12}$CO.
Next, in figure \ref{fig:3}b we find an offset between the $^{12}$CO 
and the $^{13}$CO arms.
This offset seems to be obvious, especially at the beginning of the 
spiral arm and the peak at (R.A., Decl.) = (\timeform{13h29m57s}, 
\timeform{47D10'50''}). 
Taking account of the motion of gas in the frame of galactic rotation
on the assumption of the existence of a trailing arm, the $^{13}$CO arm 
is located on 
the downstream side of the $^{12}$CO arm.
On the other hand, in figure \ref{fig:3}c which is the $^{13}$CO map 
superposed on the H$\alpha$ emission contour, we can see that the
depression on the $^{13}$CO arm is located in a region where 
the H$\alpha$ emission is not detected.
Also, the H$\alpha$ arm is located on the downstream side of 
the $^{12}$CO arm and the $^{13}$CO arm is located along and closer 
to the H$\alpha$ arm than the $^{12}$CO arm.
These observations indicate that the $^{13}$CO arm shows a closer 
correspondence to  
the H$\alpha$ emission than the $^{12}$CO emission in the 
arm region.
Figure \ref{fig:3_2} helps to clarify these features. 
This figure is an 
$R$ -- $\theta$ plot of $^{13}$CO emission superposed on ${12}$CO and
H$\alpha$ emissions, where $R$ and $\theta$  
are the distance from the center 
and the azimuthal angle for a face-on view, respectively.
These figures have a range at from $R$ = \timeform{30''} to \timeform{100''},
covering the spiral arm and not including the central region.
It makes the existence of the spiral arm clearer.
In the figures, we can see the depressions both on the $^{13}$CO 
and the H$\alpha$ spiral arms at $R$ = \timeform{50''}
while there is no prominent depression on the $^{12}$CO arm.
We can also see that the $^{13}$CO arm shows good agreement with 
the H$\alpha$ emission at $R$ = \timeform{30''} -- \timeform{50''}
and \timeform{70''} -- \timeform{80''}
while they show offsets from the $^{12}$CO arm,
or rather, they are located on the downstream side of the $^{12}$CO arm.
On the other hand, at $R$ = \timeform{60''} -- \timeform{70''},
the $^{13}$CO has an offset from both the $^{12}$CO and the H$\alpha$.
This prompts us to ask what is the difference between them?
The H$\alpha$ emission at $R$ = \timeform{60''} -- \timeform{70''}
is located at R.A. = \timeform{13h29m52s} and Decl. = \timeform{47D10'40''}.
It is clear that the distance of this H$\alpha$ emission from 
the dust lane and the $^{12}$CO arm is 
more than twice the distance compared with the other H$\alpha$ emission
(cf. figure3 in \cite{key-Rand92}). 
Therefore, we can say that this region is not located on 
the spiral arm, or rather that it has already escaped from the arm.
This exceptional region is discussed further in subsection 3.4.
Therefore, we can say that the distribution of $^{13}$CO is 
more similar to H$\alpha$ than is the $^{12}$CO on the spiral arm,
except for a part of the arm. 
We conclude that both the $^{13}$CO and the H$\alpha$ structures are 
located on the downstream side of the $^{12}$CO structure.

\subsubsection{Radial distribution}

\begin{figure}
  \begin{center}
    \FigureFile(70mm,70mm){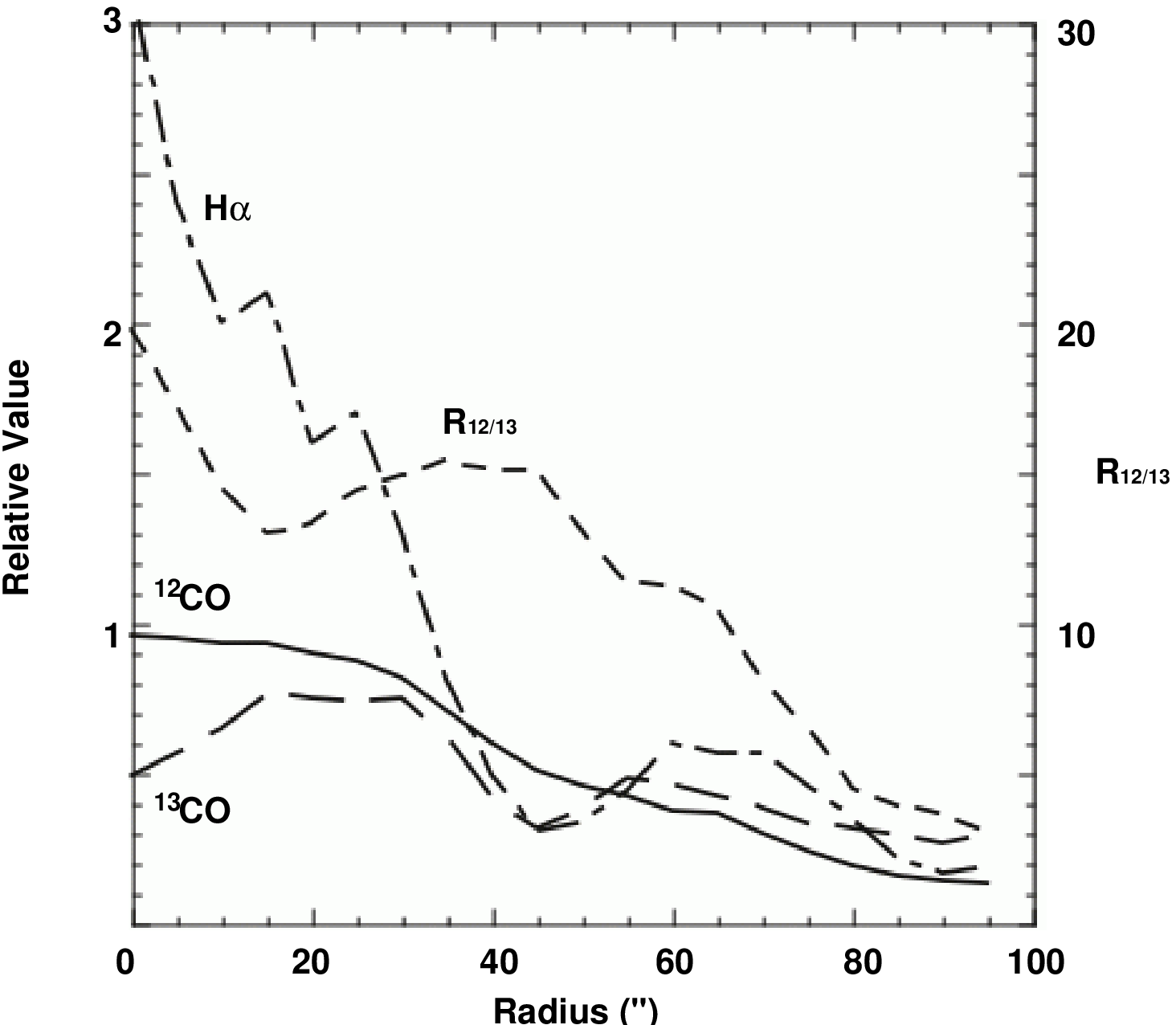}
  \end{center}
  \caption{Radial distributions of the $^{13}$CO emission, the $^{12}$CO 
   emission, the intensity ratio ($=R_{12/13}$), and the H$\alpha$ emission. 
   Those of the $^{12}$CO and the H$\alpha$ are based on the same region of 
   the observed one in $^{13}$CO.}\label{fig:4}
\end{figure}

Figure \ref{fig:4} shows the radial distribution of the $^{13}$CO emission, 
along with those of the $^{12}$CO emission and the H$\alpha$ emission.
This figure indicates the central depression and the ring-like structure
of the $^{13}$CO emission at a radius of \timeform{20''} 
-- \timeform{30''}, corresponding to 0.9 -- 1.4 kpc.
This ring-like structure reflects the central depression and
the concentrations at the bar-ends, namely the beginnings of the spiral arm.
This ring-like structure is very different from the distribution 
of the $^{12}$CO, which gradually decreases toward the outside. 
Also, the $^{13}$CO emission shows a rise at a radius of \timeform{55''}, 
which is coincident with the location of the spiral arm,  
while the $^{12}$CO gradually decreases toward the outside
of the galactic disk.
These structures of $^{13}$CO are in excellent correspondence with  
those of the H$\alpha$ emission, except for the galactic center.
On the other hand, the radial distribution of the $^{12}$CO emission 
is different from that of the H$\alpha$ emission. 

The previous studies for the kinematics of M 51 reported the positions of 
the resonances. 
These studies give the positions of the Inner Lindblad Resonance (ILR), 
which range from \timeform{20''} to 
\timeform{30''}, and that of the 4/1 resonance of \timeform{50''} -- 
\timeform{60''} 
 (\cite{key-Tully74}; \cite{key-GB93b}b;\cite{key-Aal99}).
In the radial distribution shown in figure \ref{fig:4}, 
we find two peaks at radii of
\timeform{20''} and \timeform{50''}, as mentioned above.
We suggest that the inner ring corresponds to the ILR.
Also, the outer peak is located on the radii of the 4/1 resonance.
The $^{12}$CO shows only a gradual decrease toward an outside, and does not
show peak structures at the radii of the ILR and the 4/1 resonance.

\subsection{$^{12}$CO/$^{13}$CO Ratio; Molecular Cloud Properties}

In figure \ref{fig:4}, the radial distribution of the ratio of the  
$^{12}$CO-to-$^{13}$CO integrated intensities  (=$R_{12/13}$) is also shown 
along with those of the $^{12}$CO, $^{13}$CO, and H$\alpha$ emissions.
This figure indicates that the ratio has a range from 5 to 20, and 
that the central region shows a high ratio of $\geq$ 20.
This high ratio in the central region is due to the depression of 
 $^{13}$CO in the central region.
Although the radial distribution of the $R_{12/13}$ gradually decreases 
with the radius, 
there is a slight increase between the inner ring of the $^{13}$CO and 
the outer ring or the spiral arm.
This suggests the possibility that the $R_{12/13}$ is high between 
the ring and the spiral arm, namely, in the interarm. 

\begin{figure}
  \begin{center}
    \FigureFile(100mm,100mm){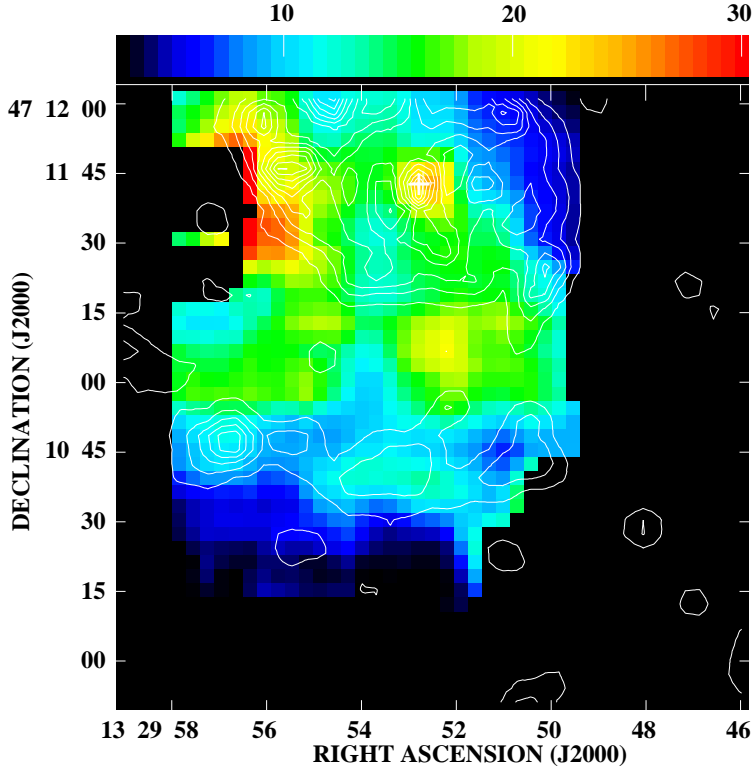}
  \end{center}
  \caption{Map of the intensity ratio ( $= R_{12/13}$ color scale)
   superposed on the H$\alpha$ emission (contour). 
   The cross indicates the galactic center of M 51.}\label{fig:5}
\end{figure}

\begin{figure}
  \begin{center}
    \FigureFile(100mm,100mm){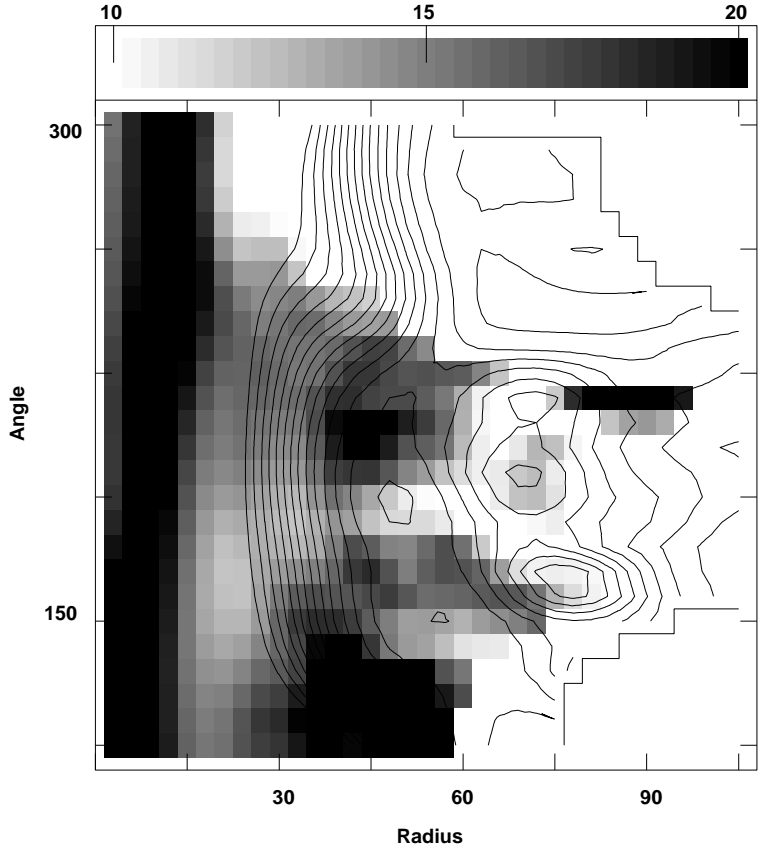} 
\end{center}
  \caption{$R$ -- $\theta$ plot of the $R_{12/13}$ (grey scale) 
  superposed on the H$\alpha$ emission (contour).
  The H$\alpha$ emission is convolved with the \timeform{17''} beam.}
\label{fig:5_2}
\end{figure}

The spatial distribution of $R_{12/13}$ is clearer in figure \ref{fig:5}
and figure \ref{fig:5_2}, 
which are the superposed color and grey scale images of the ratio  
on the H$\alpha$ emission contour.
As can be seen in these figures, $R_{12/13}$ is found to vary spatially. 
These diagrams indicate that the low $R_{12/13}$ of typically $\sim$ 10 
can be seen at the position of the spiral arm, except for a part of the region, 
and high $R_{12/13}$, $\geq$20, at the positions of most of the interarm and 
the central regions. 
We must also note that the high $R_{12/13}$ regions have an excellent 
anti-correlation with the H$\alpha$ emission except for the galactic center.
In other words, the regions with high $R_{12/13}$ such as the interarm 
show little or no H$\alpha$ emission,
while the low $R_{12/13}$ is seen in the arm region where the H$\alpha$ 
emission are distributed.
However, it must be noted that the figure shows that
there is a part of the arm with high $R_{12/13}$ and weak H$\alpha$
emission, 
which are 
located at radii from \timeform{30''} to \timeform{40''}.
We add that there are local peaks on the spiral arm.
However, their contrasts have been significantly lower than 
the arm-interarm contrast.
Only the galactic center has both high $R_{12/13}$ and strong 
H$\alpha$ emission, which is from the AGN of M 51.

We summarize the $R_{12/13}$ in various regions of our Galaxy
and galaxies in table 2.
A high $R_{12/13}$ has been seen in both IR-bright galaxies
(\cite{key-Aal95})
and starburst galaxies (e.g., M 82: \cite{key-Kiku98}), and 
in diffuse cold clouds, such as high-latitude clouds 
(\cite{key-Knap88}). 
On the other hand, the disk of our Galaxy shows a low $R_{12/13}$ value 
(\cite{key-Sol79}; \cite{key-Pol88})
and the non-active spirals, such as normal and poststarburst galaxies, 
also have intermediate ratios, $\sim10$, in the disks 
(e.g., NGC 891:  \cite{key-Saka97}).
$R_{12/13}$ in these non-active galaxies is lower than 
those of the starburst galaxies and the high latitude clouds, 
and higher than that of the disk of our Galaxy.
Our results show that the $~R_{12/13}$ values in the central region and 
the interarm of M 51 are higher than that in the spiral arm 
and are closer to those of starburst galaxies 
and high-latitude clouds.
It is also indicated that in the arm of M 51 the $R_{12/13}$ value is 
slightly higher 
than those of non-active galaxies.  
This is because $R_{12/13}$ measured on the arm of M 51 may 
include a part of that in the interarm,
which has a high $R_{12/13}$ value,
due to the narrow arm width ($\sim$ \timeform{10''}),
which was obtained by previous interferometric observations
with a higher spatial resolution (e.g., \cite{key-Rand90}).
Consequently, it is reasonable to expect that the obtained $R_{12/13}$ 
on the arm 
may be higher than that in the arm only, and is possibly similar
to that expected from the normal galaxies.

\begin{table}
  \caption{$R_{12/13}$ in galaxies.}\label{tab:2}
  \begin{center}
    \begin{tabular}{llcl}
 \hline\hline
galaxy & region & $R_{12/13}$ & Remark\\
\hline \\
Our Galaxy & Disk & 5.5 -- 6.7$^{a,b}$ \\
           & High latitude clouds & 40 $^{c}$\\
\\ 
Non-active galaxies \\
NGC 891  & Center         & 15.4$^{d}$            & Edge-on \\
        & Total          & $6.6^{d}$\\
NGC 1808 &                & 7.8$^e$               & Hot spots \\
NGC 4945 &                & 10 -- 19$^f$             & Edge-on \\
NGC 7331 & Center         & 6.7 -- 8.8$^g$           & Poststarburst \\
\\ 
Active galaxies \\
NGC 7479 & Bar            & 5 -- $\geq$30$^{h}$ & Starburst\\
M 82     & Central region & 20 -- $\geq$70$^{i}$ & Starburst \\
NGC 4194 &                & 20$^j$                & Medusa merger \\
Arp 299 & IC694 nucleus  & 60$^j$                & Merger \\
        & IC694 disk     & 10$^j$                & \\
        & NGC3690        & 13$^j$   & \\
IR bright galaxies & center     & 5 -- 35$^{k}$\\ 
\\ 

M 51 & Center             & 8 -- 9$^{l}$               & \\
    &                    & $\geq$ 20$^{m}$       & \\
    & Arm                & 9 -- 10$^{l}$  \\
    &                    & 10$^{m}$ \\
    & Interarm           & $\geq 15^{l}$ \\
    &                    & 20$^{m}$ \\ 
\hline

{\footnotesize a): \citet{key-Sol79}}. \\
{\footnotesize b): \citet{key-Pol88}}. \\
{\footnotesize c): \citet{key-Knap88}}. \\
{\footnotesize d): \citet{key-Saka97}}. \\
{\footnotesize e): \citet{key-Aalt94}}. \\
{\footnotesize f): \citet{key-Berg92}}. \\
{\footnotesize g): \citet{key-Isr99}}. \\
{\footnotesize h): \citet{key-Hutt00}}. \\
{\footnotesize i): \citet{key-Kiku98}}. \\
{\footnotesize j): \citet{key-Aal97}}. \\
{\footnotesize k): \citet{key-Aal95}}. \\
{\footnotesize l): \authorcite{key-GB93a}(1993a)}. \\
{\footnotesize m): This work} \\
\end{tabular}
\end{center}
\end{table}

\begin{figure}
  \begin{center}
    \FigureFile(100mm,100mm){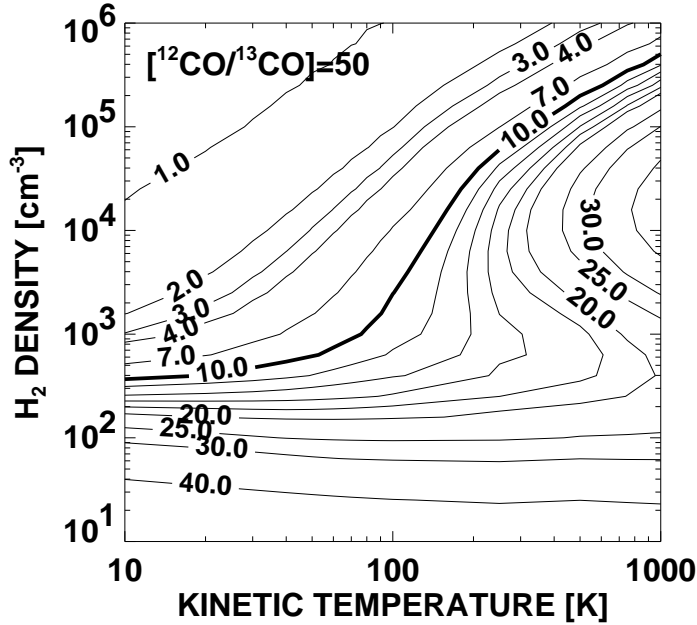}
  \end{center}
  \caption{Density (vertical axis) and temperature (horizontal axis)
   dependence of the $^{12}$CO/$^{13}$CO intensity ratio ($=R_{12/13}$;
   \cite{key-Matsu00}).
   The solid contours show curves of constant $R_{12/13}$,
   and the thick solid curve indicates $R_{12/13}$ of 10.0.}\label{fig:7}
\end{figure}

In order to understand the physical properties of the molecular gas
in the central and disk regions of M 51 (and also other galaxies) from
the $^{12}$CO and $^{13}$CO data, we used the Large-Velocity-Gradient (LVG)
calculations \citep{key-Gol74,key-Sco74} assuming a one-component model.
$R_{12/13}$ was calculated as a function of the H$_{2}$ number density
from $10^{1}$ to $10^{6}$ cm$^{-3}$, and the kinetic temperature from 10 to
1000 K.
The collision rates for CO molecules are available from
\citet{key-Flo85} ($\leqq250$ K) and \citet{key-Mck82} ($\geqq500$ K).
We fixed the molecular abundances to the `standard' relative abundance as
$Z(^{13}{\rm CO})$ = [$^{13}$CO]/[H$_{2}$] $=1\times 10^{-6}$ \citep{key-Sol79}
and [$^{12}$CO]/[$^{13}$CO] = 50.
We also fixed the velocity gradient, $dv/dr$, as a general value of 1.0
km s$^{-1}$pc$^{-1}$.
The calculated results are shown in figure~\ref{fig:7} \citep{key-Matsu00}.
This diagram indicates that $R_{12/13}$ depends mainly on density
when the temperature is not high ($<$100 K).
In the Galactic disk, the temperature of typical molecular clouds has 
a typical value of 10 K.  
In this case, it seems reasonable that $R_{12/13}$ is a good tracer 
for density of molecular clouds, namely $R_{12/13}$ decreases
with the density.
Our results show that a low $R_{12/13}$ value, smaller than 10, is seen in 
the arm associated with the H$\alpha$ emission, and 
the interarm and a part of the arm with no associated H$\alpha$ 
emission have a high $R_{12/13}$ value of $\sim$ 20, 
as mentioned above.
This figure indicates that there are diffuse molecular clouds with a density
of 2$\times 10^2$ cm$^{-3}$ in the interarm and the arm with 
no associated H$\alpha$ emission,
whereas most molecular clouds in the arm associated 
with the H$\alpha$ emission  
have a density larger by a factor of two,  
in the case that the same temperatures of molecular clouds in each region,
e.g., 10 K, is assumed.  
Namely, the massive star-forming regions traced by the H$\alpha$ emission
are located in the denser clouds with a low $R_{12/13}$ value.

On the other hand, the galactic center as well as the interarm 
region also has a high $R_{12/13}$ value of 
$\geq$ 20, 
which is not consistent with the results obtained by 
\authorcite{key-GB93a}(1993a).
The reason for this may be the difference in the spatial resolution
between their (\timeform{23''}) and our (\timeform{17''}) observations.
Since our $R_{12/13}$ map shows a very low ratio 
in the bar-end/beginning of the spiral arm, the larger beam 
can cause contamination with this low-ratio region.
To confirm this, we convolved our maps to their resolution and
measured the ratio at the center. 
The measured ratio is 12 -- 13, which is in agreement with their value
(8 -- 9) within the errors.
As a result, it is reasonable to expect that $R_{12/13}$ decreases 
in their study.

Figure \ref{fig:7} shows that the high $R_{12/13}$ values at the central
region of M 51 suggests two possibilities for the physical condition
of molecular gas: one is a low-density ($\sim10^{2}$ cm$^{-3}$)
condition; the other is a high-density ($10^{4\pm1}$ cm$^{-3}$)
and high temperature ($\geq$ 300 K) condition.
To define the physical conditions of the molecular gas at the center,
we compared our results with the previously published high resolution
observations.
The distribution of molecular gas at the central region of M 51 is
strongly concentrated toward the center; interferometric $^{12}$CO
\citep{key-Saka98,key-Aal99} and HCN \citep{key-Kohno96} observations
clearly show a centrally peaked distribution, and suggest disk-like
kinematics surrounding the low-luminosity AGN
(see also \cite{key-Sco98}).
On the other hand, the high-resolution $^{13}$CO observations
display a lack of $^{13}$CO emission toward the center
\citep{key-Matsu98}.
The HCN/$^{12}$CO intensity ratio suggests that the central region
is rich in dense molecular gas with a density of
$\sim10^{4-5}$cm$^{-3}$ \citep{key-Kohno96}.
The HCN/$^{13}$CO ratio shows a high value ($>3$) and suggests that
the central region is dominated by hot ($\geq$ 100 K) and dense
($\sim10^{5\pm1}$ cm$^{-3}$) gas \citep{key-Matsu98}, and may be
correlated with the AGN \citep{key-Matsu99}.
If our observations are to be consistent
with previous ones, and if the density is roughly $10^5$ cm$^{-3}$, 
then our value of $R_{12/13}$ (20) implies a temperature of 
several hundred K.
Active galaxies, such as starbursts, also have high $R_{12/13}$ values
as mentioned above, which may also be due to the molecular clouds
with high density and high temperature.
We can say that the central region of M 51 has hot, dense clouds
as do active galaxies.

\subsection{Kinematics}

\begin{figure}
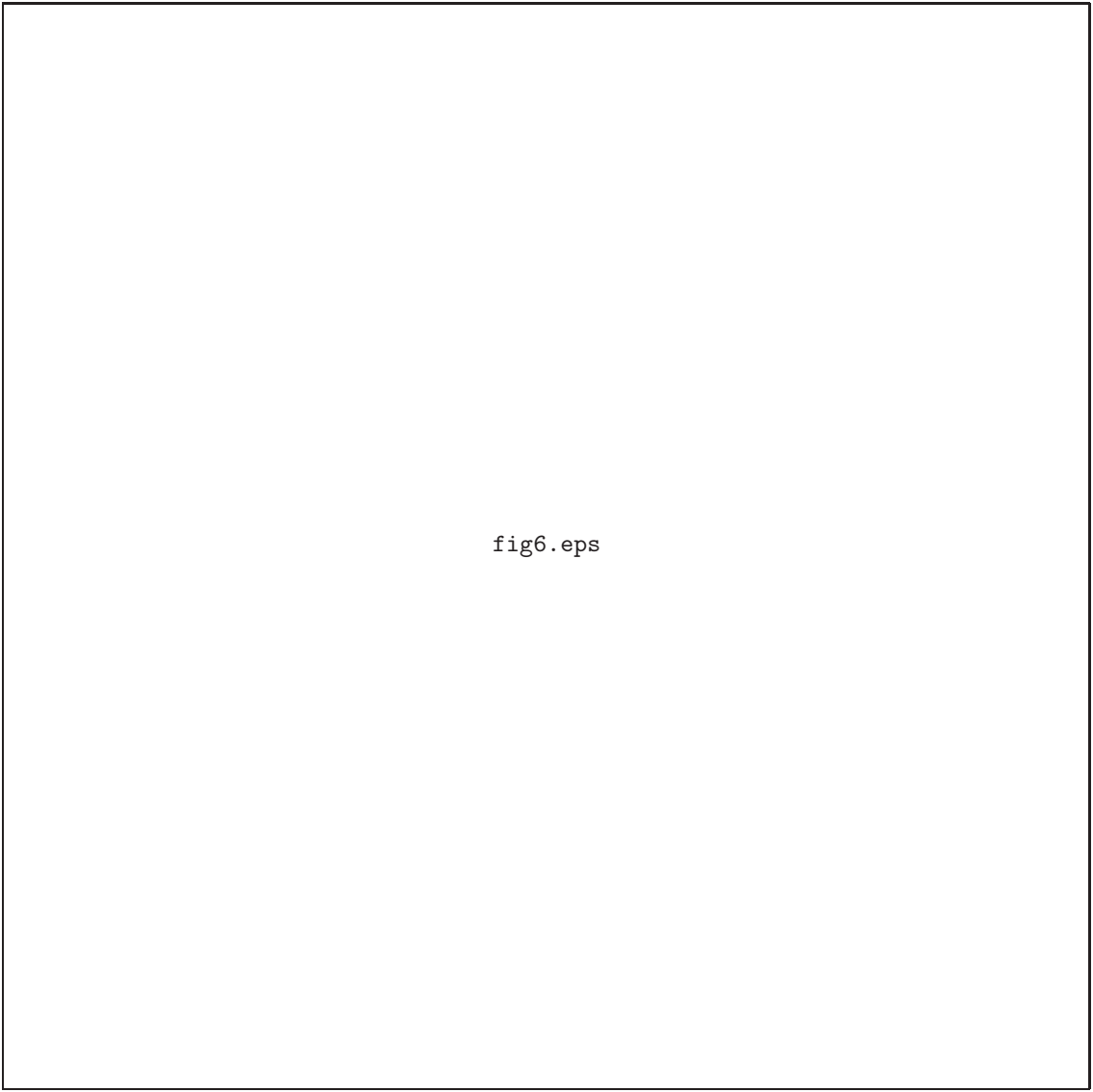

  \begin{center}
    \FigureFile(150mm,150mm){fig6.eps}
  \end{center}
  \caption{(a) Velocity field. The contour interval is 10 km s$^{-1}$ and 
      the thick line indicates 460 km s$^{-1}$.
      The dashed lines indicate the major and minor axes.
  (b) P--V diagrams along major axis. 
     The arrow indicates the location of the arm. 
     The contour interval is 15 mK and the lowest contour is the 30 mK.}
\label{fig:6}
\end{figure}

Figure \ref{fig:6}a shows the velocity field of the $^{13}$CO emission,
which was made by tracing the peak velocity of the $^{13}$CO line profile.
The previous observations of H~{\sc I},$^{12}$CO and H$\alpha$ emissions for 
the large area provide a position angle of M 51 of $170^\circ$ 
(\cite{key-Tully74}).
In the central region of figure \ref{fig:6}a, we found a 
feature difference from that expected from the position angle, 
P.A. = 170$^\circ$, 
namely isovelocity contours show a tilt from the minor axis of the 
galaxy.
It is reported from near-infrared observations 
that there is an oval potential in the central region of M 51 
(\cite{key-Pie86}; \cite{key-TG88}).
An oval potential causes a distortion of the velocity field; 
this feature in M 51 was found in previous
observations (e.g., \cite{key-Kuno97}).

The isovelocity contours also show S-shape disturbances 
from pure rotation at the spiral arm. 
This is consistent with previous studies of $^{12}$CO.
This could be due to streaming motion caused by a density wave,
as was already suggested based on previous studies with $^{12}$CO observations
(\cite{key-GB93a}a; \cite{key-Kuno97}).

Because the observations give only the velocity component
in the line of sight directly, we present a Position--Velocity (P--V) diagram 
along the major axis (P.A. = $170^\circ$) to indicate the tangential velocity
(figure \ref{fig:6}b).
The width of the strip used to make the P--V diagram along the major axis
is \timeform{17''}. 
We find that there is a central depression and 
a concentration at the end of the rigid rotation
($R \sim$ \timeform{20''}) in figure \ref{fig:6}b.
We can also see a velocity shift of 15 km s$^{-1}$
at the position of the spiral arm indicated by  arrow in the figure. 
These correspond to $\sim$ 40 -- 50 km s$^{-1}$
on the plane of the galaxy.
This was pointed out by previous $^{12}$CO observations 
(\cite{key-Kuno97}; \cite{key-Aal99}),
and it was suggested that molecular gas is shocked across the spiral arm. 
This figure suggests that the dense gas which is traced by $^{13}$CO
as well as diffuse gas traced by $^{12}$CO is shocked by the density wave.

\subsection{Time Delay of Dense Gas Formation and Star Formation}

As mentioned in section 3.2, the $^{13}$CO emission shows good 
agreement with the H$\alpha$ emission rather than with the $^{12}$CO one,
except for in a part of the arms, and they are located on 
the downstream side of the $^{12}$CO.
Here, we discuss the effect of the spiral arms, namely the density wave
on the dense gas and the star-formation mechanism.
Before turning to this problem, we must consider
the exceptional H$\alpha$ emission region, which shows an offset from 
the $^{13}$CO emission.

Because this region is far from the dust lane and the $^{12}$CO arm,
as mentioned in subsection 3.1.2, it seems reasonable to expect
that it has escaped from the spiral arm.
If the star formation is triggered by a shock originating from the 
density wave,
it indicates that this region is an ``older, or a more evolved star 
forming region'' than those on the spiral arm, 
since the location of the shock is supposed to be the dust lane.
We may say that, in the evolved star-forming region, 
the dense gas decreases due to consumption by the star formation, 
the ionization, the photo-dissociation and so on, 
and the dense gas formation, itself, is suppressed because of 
a decrease in the surrounding gas when it is far from the arm.   
For this reason, it seems reasonable that the above 
H$\alpha$ emission region has no associated dense gas 
which is traced by $^{13}$CO. 
Therefore, when we consider the effect of the density wave on
the formation mechanism of the dense gas and the following star,
it should be an exception.

\begin{figure}
  \begin{center}
    \FigureFile(80mm,150mm){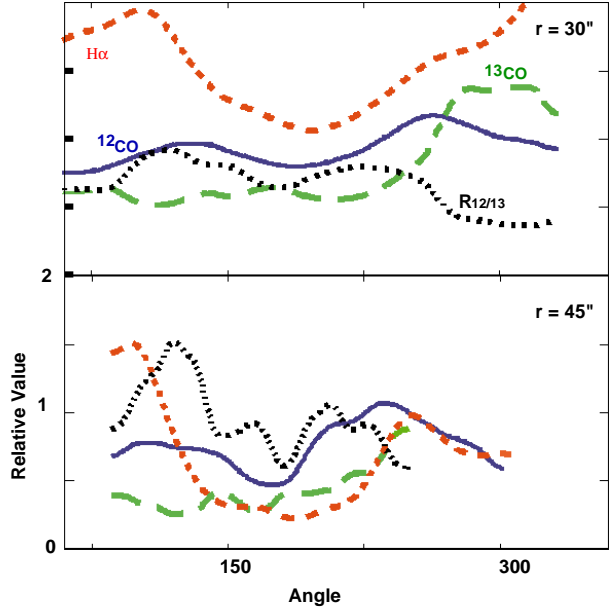}
  \end{center}
  \caption{Azimuthal distributions of the $^{13}$CO emission, the 
   $^{12}$CO emission and the intensity ratio obtained with NRO-45m, 
   and the H$\alpha$ emission convolved to the \timeform{17''} beam 
   at radii of  \timeform{30''} and \timeform{45''}.
   The horizontal and vertical axes of each panel are the azimuthal angles
   (in degree) and their relative values, respectively. The gas flows from
   left to right in this figure.}\label{fig:8}
\end{figure}

In order to investigate the effect of a spiral arm on the nature of 
molecular gas, we present 
the azimuthal distributions of $^{12}$CO, $^{13}$CO,
$R_{12/13}$, and H$\alpha$ at radii of   
\timeform{30''} and \timeform{45''} (figure \ref{fig:8}).  
Although it indicates that the $^{12}$CO and $^{13}$CO emissions increase 
toward the spiral arm from the interarm, their peaks are not coincident,
namely, the rise of the $^{13}$CO is delayed compared with that of 
the $^{12}$CO.
As a result, $R_{12/13}$ also shows a decrease across the spiral arm 
and the trend mentioned above is confirmed.  
The delays are 10 -- 20 deg at radii from \timeform{30''} 
to \timeform{50''} and are smaller at larger radii.
This suggests that there is a time delay of the formation of 
dense gas traced by $^{13}$CO from the accumulation of molecular gas
by the density wave. 
The small delay (in the unit of degree) at a large radius
is reasonable, if there is the same time delay between the gas accumulation 
caused by the density wave and the dense gas formation. 
The difference, $d$, is 400 -- 800 pc at a radius of \timeform{48''},
corresponding to 2.2 kpc.
We can estimate the time delay corresponding to $d$ from
the rotation curve.
This time delay is given by $\tau_{\rm d}$ = $d/V_{\rm arm}$, where
$V_{\rm arm}$
is the velocity perpendicular to the arm in the frame corotating 
with the spiral pattern.
For the adopted pattern speed of $\Omega_{\rm p}$ = 14 km s$^{-1}$ kpc$^{-1}$
(see \cite{key-Kuno95}) and the pitch angle of 20 degrees, 
we can obtain the value of $V_{\rm arm}$ = 58 km s$^{-1}$.
This is the upper limit of the $V_{\rm arm}$ because we assume 
a circular motion.
Then the corresponding time delay is $\tau = (0.7 - 1.5) \times10 ^7$ yr
at 2.2 kpc.

We must note that the estimated time delay depends on the adopted 
pattern speed. 
That is, if a larger pattern speed is adopted, $V_{\rm arm}$ becomes
small and  $\tau$ becomes large.
Actually, the pattern speed obtained by previous studies also has a result of
27 km s$^{-1}$ kpc$^{-1}$ (\cite{key-GB93b}b), and may be larger. 
In the case of larger pattern speed, the above estimate of 
the time delay is the lower limit.
We add that $\tau$ does not increase 1.2 times even if the pattern
speed doubles.

It has been known that the H$\alpha$ emission, indicating massive star 
formation, is also located on the downstream side of the $^{12}$CO emission
(\cite{key-Vog88}; \cite{key-Rand92}).
As noted above, it shows good correspondence of 
the $^{13}$CO with the H$\alpha$ in the disk of M 51.
From these results, we presume that the star formation occurred 
following dense gas formation later than $\sim 10^7$ yr 
after the accumulation of diffuse molecular gas caused by density wave.

Here, in order to address the origin of the delay of the dense gas and 
the star formation from the compression of the gas due to the density wave,
we consider the gravitational instability of molecular gas
in the spiral arm. 
Here, we can introduce the Toomre $Q$ parameter to estimate 
the effect of gravitational instability on the mechanism for 
dense gas formation (e.g. \cite{key-Ken89}).
The previous studies suggest that the $Q$ parameters have a correlation with 
the star-formation activity (\cite{key-Kohno01}; \cite{key-Shio98}; 
\cite{key-Tosa97}; \cite{key-Ken89}). 
The $Q$ parameter is expressed as $Q = \Sigma_{\rm crit}/\Sigma_{\rm gas}$,
where $\Sigma_{\rm gas}$ is the gas surface density and $\Sigma_{\rm crit}$
is critical density for gravitational instabilities, and 
is used as a criteria for local instability in an isothermal thin disk.
$\Sigma_{\rm crit}$ is given by

\begin{equation}
\Sigma_{\rm crit} = \alpha \frac{\sigma_{\rm v} \kappa}{\pi G}
 = 74 \alpha \left(\frac{\sigma_v}{{\rm km~ s}^{-1}}\right) \left(\frac{\kappa}{{\rm km~ s}^{-1} {\rm pc}^{-1}}\right) M_\odot {\rm pc}^{-2}, 
\end{equation}
where $\sigma_{\rm v}$ and $\kappa$ are the velocity dispersion
and the epicyclic frequency for the gas disk, respectively. 
$\alpha$ is a dimensionless constant (\cite{key-Toom64}).
$\kappa$ is expressed as

\begin{equation}
\kappa = \left\{ 2\frac{V(r)}{r} \left[\frac{V(r)}{r} + \frac{dV(r)}{dr}\right] \right\} ^{0.5}
\end{equation}
where $V(r)$ is rotational velocity at the radius of $r$.
In the case that we use 10 km s$^{-1}$ as the velocity dispersion and 
0.63 as  $\alpha$ (\cite{key-Ken89}),
we obtain a $\Sigma_{\rm crit}$ of $84 M_\odot$pc$^{-2}$
and 55 $M_\odot$pc$^{-2}$
at the radii of \timeform{30''} and \timeform{50''}, respectively.
It should be added that we have used the observational rotation curve,
which was fitted with the potential model of \citet{key-Miya75},
as the rotational velocity and the gradient (see \cite{key-Kuno97}).  
Because the surface density of the gas at the spiral arm is 
$\sim 200~ M_\odot$pc$^{-2}$ (\cite{key-Kuno95}), 
the $Q$ values are 0.4 and 0.3.
This indicates that the gas in the arm is gravitationally unstable.
Therefore, dense gas formation in the arm may occur 
via a gravitational instability of the gas;
consequently the star formation occurs there.

Next, we compare the above time scale with those of gravitational
instabilities.
The growth time for a gravitational instability can be estimated
as follows (e.g., \cite{key-Lar87}):

\begin{equation}
\tau = \frac{\sigma_{\rm v}}{\pi G \Sigma_{\rm gas}},
\end{equation}
where $\sigma_{\rm v}$ and $\Sigma_{\rm gas}$ are the velocity dispersion and
the average surface density of gas, respectively.
We can adopt these values at a radius of \timeform{40''} as 
10 km s$^{-1}$ and 200 $M_\odot {\rm pc}^{-2}$, respectively,
from the peak of the $^{12}$CO arm (\cite{key-Kuno95}).
The estimated timescale is $\sim 10^7$ years, which is similar to
the above time delay between the $^{12}$CO and the $^{13}$CO, and 
we can explain the mechanism of dense gas formation by
the gravitational instability.

We must note that an offset between the $^{12}$CO and the HCN has been found
in the central region of NGC6951 (\cite{key-Kohno99}), which 
corresponds to $\sim 10^6$ yr.
HCN is a tracer of dense gas with a density of 10$^{4-5}~{\rm cm}^{-3}$, 
which is higher than $^{13}$CO.
This value is also similar to the timescale of gravitational instabilities.
It seems reasonable to suppose that dense gas formation 
occurs after $\sim 10^{6-7}$ yr from the accumulation of gas.
As a result, star formation also occurs after that.
It is likely that gravitational instability plays 
an important role in the mechanism of dense gas and star formation.

\section{Conclusions}

We present the results of $^{13}$CO($J$~=~1--0) 
mapping observations toward
the southern bright arm region of nearby spiral galaxy M 51 
carried out with the Nobeyama 45 m telescope. 
The main conclusions are summarized as follows:

1. We find a depression of the $^{13}$CO emission in the central
\timeform{20''} -- \timeform{30''} region, corresponding to $\sim$1 kpc. 
There is a ring-like structure surrounding the central region,
which corresponds to the bar-end or the beginning of the spiral arm.
 
2. The arm structure in the $^{13}$CO is shown in the total integrated 
intensity map, 
and the arm-to-interarm ratio is 2 -- 4.  
The ratio of the $^{13}$CO in the observed region is higher 
than that of the $^{12}$CO.
Although it shows a global correspondence with the $^{12}$CO, 
a detailed comparison shows spatial differences between them, 
e.g., there is a depression of the $^{13}$CO on the spiral arm.

3. The $^{13}$CO distribution shows a good correspondence with that of 
the H$\alpha$ emission, except for in the galactic center,
rather than the $^{12}$CO. 
The $^{13}$CO and the H$\alpha$ show the depression 
at the same position on the spiral arm and are located on the downstream 
side of the $^{12}$CO.

4. A velocity shift is detected at the position of the spiral arm,
which suggests a streaming motion caused by the density wave.

5. The $^{12}$CO/$^{13}$CO ratio has a range from 4 to $\geq$20.
The central region and the interarm indicate a high value of $\sim$20,
while a low value of $\sim$10 is shown in the arm.
They indicate that there is denser gas in the arm than in the interarm.
The densities in the arm and the interarm regions were derived to be
$\geq 6\times 10^2$ and $2\times 10^2$ cm$^{-3}$, respectively, 
based on the LVG calculation.  
The high ratio in the central region is due to very hot and dense gas 
related to AGN in the nucleus of M 51.   

6. The azimuthal distribution at a constant radius shows that
the $^{13}$CO emissions are located on the downstream side of
the $^{12}$CO emission. 
This indicates that there is a time delay from the accumulation of gas
caused by shock to the dense gas formation, and the resultant star formation.
This time delay leads to $\sim 10^7$ yr from the
galactic rotation. 
This timescale is similar to the growth time for the gravitational instability.
It suggests that gravitational instability plays an important role
in dense gas formation.

\vspace{0.5cm}

We are grateful to NRO staff for the operation and the 
improvement of the 45-m telescope.
We would like to thank Dr.\ K. Kohno for useful comment and discussion.
We appreciate the encouragement, comments and discussion with Dr.\ H. Okuda, 
especially concerning the narrow-band observation at GAO and we also thank to 
Mr.\ H. Kawakita for observations of GAO 65-cm.  
Y.S. thanks the Japan Society for Promotion of Science (JSPS)
Research Fellowships for Young Scientists. 
S.M. thanks the JSPS Fellowships for Research Abroad.




\end{document}